\pdfoutput=1 
\documentclass[11pt,a4paper]{jinst} 
\let\ifpdf\relax
\usepackage{amsmath}
\usepackage{multirow}
\usepackage{lineno}
\usepackage{subfigure}
\usepackage{atlasphysics} % Contains useful shortcuts. Uncomment to use
\usepackage[update,prepend]{epstopdf}
\title{A neural network clustering algorithm for the ATLAS silicon pixel detector}

\author{The ATLAS Collaboration}

\date{\today}

\abstract{A novel technique to identify and split clusters created by multiple charged particles in the ATLAS pixel detector using a set of artificial neural networks is presented. Such merged clusters are a common feature of tracks originating from highly energetic objects, such as jets. Neural networks are trained using Monte Carlo samples produced with a detailed detector simulation. This technique replaces the former clustering approach based on a connected component analysis and charge interpolation. The performance of the neural network splitting technique is quantified using data from proton--proton collisions at the LHC collected by the ATLAS detector in 2011 and from Monte Carlo simulations. This technique reduces the number of clusters shared between tracks in highly energetic jets by up to a factor of three. It also provides more precise position and error estimates of the clusters in both the transverse and longitudinal impact parameter resolution.}

\begin{document}

%\ifIsINT
%\newpage
%\tableofcontents
%\newpage
%\fi 

% introduction

\section{Introduction}

Track and vertex finding is one of the most challenging tasks in reconstructing events from proton--proton collisions recorded by the ATLAS detector~\cite{Aad:2008zzm} at the Large Hadron Collider (LHC) at CERN. ATLAS is a general--purpose detector designed to make precision measurements of known physics processes and to probe new physics at the energy frontier of the LHC. At the centre of the detector is an optimised, multi--technology tracking detector embedded in a 2~TeV axial magnetic field produced by a solenoid. This central tracking detector is designed to enable the reconstruction of trajectories (also called tracks) from charged particles and is called the inner detector (ID). It has a silicon pixel detector at the innermost radii, surrounded by a silicon microstrip detector and a straw--tube detector called the transition radiation tracker (TRT) that also identifies particle types using transition radiation. The ID provides full $\phi$ coverage for particles produced within $|\eta| < 2.5$~\footnote{The ATLAS reference system is a Cartesian right--handed coordinate system, with the nominal collision point at the origin and the $z$--axis along the beam direction. The positive $x$--axis is defined as pointing from the collision point to the centre of the LHC ring and the positive $y$--axis points upwards. The azimuthal angle, $\phi$, is measured around the beam axis, and the polar angle, $\theta$, is measured with respect to the $z$--axis. The pseudorapidity is defined as $\eta = - \ln{(\tan{(\theta/2)})}$.} ($|\eta|<2.0$ in the case of the TRT) in the LHC beam interaction region. The ID is surrounded by electromagnetic and hadronic calorimeters. The outermost detector is the Muon Spectrometer (MS) which is a stand--alone tracking detector with a toroidal magnetic field. Trajectories in the MS are matched to ID trajectories to reconstruct combined muon candidates~\cite{Aad:2014zya}.

Precisely measuring the trajectories of charged particles is fundamental to many analyses. Tracks are used to reconstruct primary and secondary vertices. Primary vertices are found within the beam interaction region and provide information about the number and positions of the individual proton--proton collisions. Secondary vertices are found outside the beam interaction region and are used to identify decays of heavy--flavour and long--lived particles.

Track reconstruction algorithms use local and global pattern recognition algorithms to identify detector measurements that were produced by a charged particle. A track fitting procedure is used to estimate the compatibility of measurements with the track hypothesis under the assumptions of a track model. The term ``measurement" is used to describe clusters from the pixel detector and the SCT, and drift circles from the TRT when assigned to a track. The formation of clusters from individual hits is explained in section~\ref{sec:cluster}. The ID track reconstruction consists of several sequences with different strategies as described in ref.~\cite{Cornelissen:1020106}. The main sequence is referred to as inside--out track finding, which starts from track seeds formed through combinatorial grouping of clusters in nearby layers. Track candidates are obtained by performing a search with a combinatorial filter~\cite{Mankel1997169} for additional measurements within a road following the direction of the seed. These track candidates must satisfy rather loose requirements on the number of associated measurements, the number of missing measurements and the quality of the fit. At this stage, multiple track candidates can be found for a single particle and each measurement can be associated with several track candidates. Such multiply--assigned measurements will be referred to as shared. Track candidates are then examined by an ambiguity processor that runs on this ensemble of tracks and aims to resolve ambiguities. This suppresses fake tracks built from random measurement combinations and remove track duplicates. These occur when multiple track candidates with almost identical measurement content pass the ambiguity processing. A scoring schema that analyses the measurements assigned to the track, in particular the shared measurements and holes, is used to rank the tracks. The fit quality of the individual track candidates is taken into account when the tracks are ranked according to their achieved score. Holes are defined as active detector modules where a measurement would be expected by the track prediction but none was assigned. 

The requirement that a track candidate has a low fraction of shared measurements is a very strong handle to suppress duplicate and fake tracks. However, in signatures with many collimated particles, such as the core of very energetic jets or the decay of boosted $\tau$ leptons into multiple hadrons, shared measurements can occur because of cluster merging, when the detector granularity is insufficient to resolve close--by particles. Therefore tracks passing the ambiguity-resolving stage are allowed to have a small fraction of shared measurements~\footnote{Tracks are required to have at least six non--shared measurements and at most one shared measurement.} to avoid compromising the track reconstruction efficiency. Shared measurements typically provide a worse position estimate and hence degrade the quality of the reconstructed track. A correct modelling of shared measurements in Monte Carlo simulation is important for the control of systematic uncertainties on the track reconstruction efficiency.

In this paper, a novel technique to identify and split merged clusters in the ATLAS pixel detector by using a set of artificial neural networks (NN), or NN clustering, is introduced. This leads to a reduction of approximately a factor of three in the number of measurements shared between tracks. It also significantly improves the precision of the position estimate of the clusters and the track performance as will be discussed in section~\ref{sec:nnperf}.

The data used in this document were collected by the ATLAS experiment in 2011 with the LHC colliding protons at a centre--of--mass energy of $\sqrt{s} = 7$ TeV. The studies presented here did not use the full dataset but used sufficient data events for the statistical uncertainty to be negligible. In all cases, this required only a small fraction of the 2011 dataset: in some cases a single run or even fraction of a run. The NN clustering was introduced during the 2011 data-taking period and became the default approach for the rest of the Run--1 data taking. Data taken earlier in 2011, prior to the introduction of the NN clustering algorithm, were reprocessed to guarantee a consistent and improved dataset for analyses of the entire Run--1 dataset. The new technique will be an essential component of the ATLAS track reconstruction in the upcoming Run--2, when the ID will be equipped with a new additional innermost silicon layer, which will be located closer to the beam interaction region than the innermost layer of the pixel detector~\cite{Capeans:1291633}.

In section~\ref{sec:cluster} a description of cluster measurements in ATLAS is provided. Section~\ref{sec:pix_clus_split} explains the use of a neural network to split pixel clusters. Section~\ref{sec:nnperf} shows the performance of the neural network algorithm in simulation and data. Finally, section~\ref{sec:conclusion} presents the conclusions.

\section{Cluster measurements with the ATLAS pixel detector}
\label{sec:cluster}
Due to its proximity to the beam interaction region, the pixel system plays a special role in the estimation of the distance of closest approach of a charged particle trajectory to the interaction point, commonly referred to as the impact parameter. The impact parameter is described using transverse ($d_0$) and longitudinal ($z_0$) components.

\begin{figure}[htb!]
\begin{center}
\includegraphics[width=0.8\textwidth]{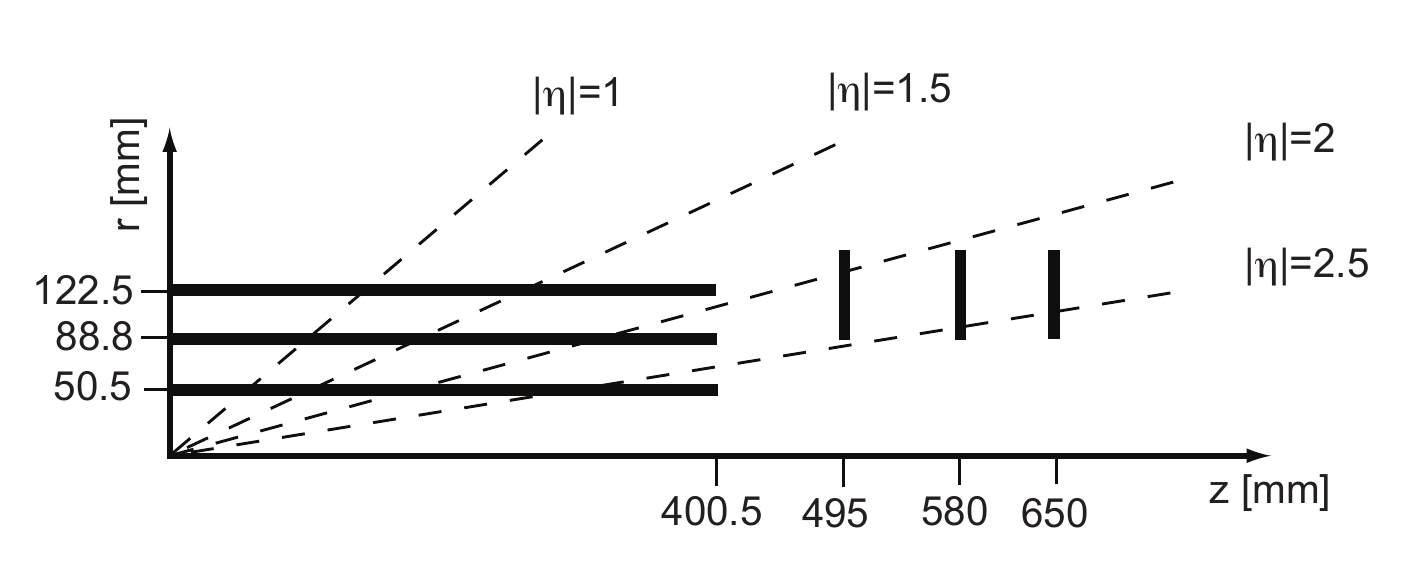}
\caption{Geometry of the ATLAS pixel detector. There are three concentric cylindrical barrel layers and three discs on each side, all are equipped with identical pixel modules.\label{fig:pixel_detector}}
\end{center}
\end{figure}

The pixel system, as shown in figure~\ref{fig:pixel_detector}, consists of three concentric cylindrical barrel layers at radii of 50.5, 88.5 and 120.5 mm around the beam pipe, and three pixel discs on each side that extend to 650 mm along the beam line to cover regions with pseudorapidity of $|\eta| < 2.7$. Cylinders and discs are constructed from a total of 1,744 identical modules, each containing 47,232 individual pixels connected to 16 front--end chips.

The modules are arranged such that the sensor segmentation defines a precise transverse measurement in the bending plane of the magnetic field and a less precise measurement in the longitudinal direction aligned with the beam line in the barrel. The modules on the discs are arranged such that the less precise measurement is in the radial direction. In the local frame of the module, the transverse direction is referred to as the local $x$ (corresponding to the pixel rows) and the longitudinal direction as the local $y$ direction (corresponding to the pixel columns). 

A typical pixel has a size of $50~\mu$m ($400~\mu$m) in the transverse (longitudinal) direction. In the region between the readout chips the longitudinal pixel size is $600~\mu$m; these are referred to as long pixels. The silicon sensor thickness is 250 $\mu$m. The modules in the barrel are shingled along the beam direction and held in a turbine arrangement in azimuthal direction at an angle of $20^{\circ}$. See ref.~\cite{1748-0221-3-07-P07007} for further details of the pixel detector.

The entire pixel system has 80.8 million readout channels. Each individual pixel is read out if the amount of charge collected in that pixel exceeds a tuneable threshold of 3500 electrons. The collected charge is obtained by measuring the pulse height using the time--over--threshold (ToT) technique. When a charged particle traverses a pixel sensor, charge is typically collected in more than one pixel. This is predominantly due to the incident angle of the particles with respect to the sensor and also the diffusion of electrons and holes in the sensor and drift of the charge generated along the track by the magnetic field. 

\begin{figure}[htb!]
\begin{center}
\includegraphics[width=0.8\textwidth]{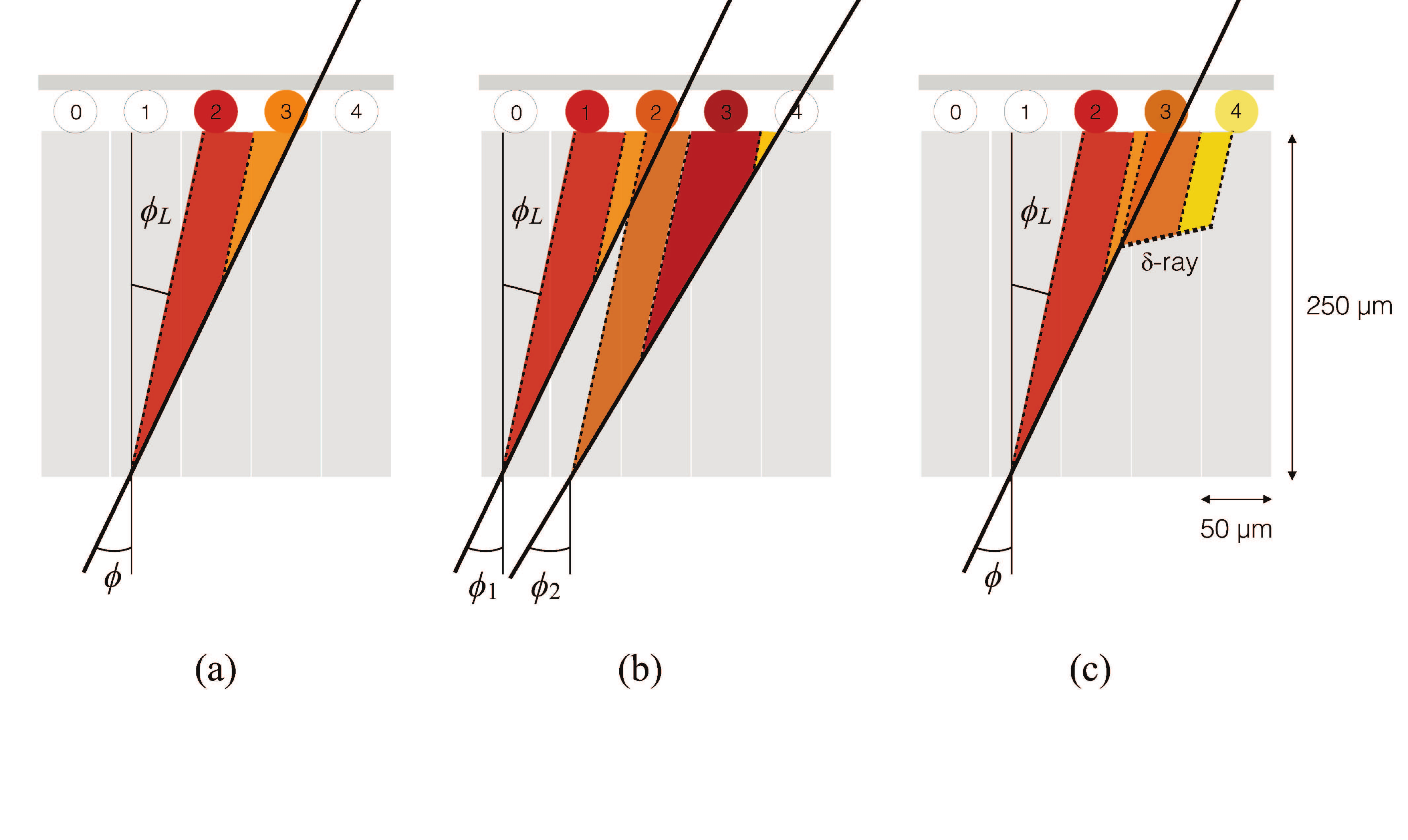}
\caption{Illustration of cluster sizes in the transverse direction for selected scenarios:
(a) a single charged particle traversing the silicon sensor with incident angle, $\phi$, producing a cluster with a width of two pixels. The charge drifts due to the electric and magnetic field in the direction of the Lorentz angle $\theta_\mathrm{L}$ to the surface.
(b) a merged cluster produced by two close-by particles ($A_1$, $A_2$) that both deposit charge in the second pixel. The fourth pixel is not read-out as the charge is below the read-out threshold.
(c) a large cluster created from a single charged particle due to a $\delta$-ray produced in the silicon\label{fig:pixel_track}}
\end{center}
\end{figure}

For primary charged particles, originating from beam collisions and with $p_\mathrm{T} > 400$~MeV, clusters vary in size between $1.4$ and $3$ pixels in the transverse direction and between 1 and approximately 3.5 pixels in the longitudinal direction. Illustrations of different scenarios of charged particles traversing a silicon sensor are shown in figure~\ref{fig:pixel_track}. In figure~\ref{fig:pixel_track} (a), a single charged particle traverses the silicon sensor with incident angle, $\phi$, and produces a cluster with a width of two pixels. The charge drifts due to the electric and magnetic field in the direction of the Lorentz angle $\theta_\mathrm{L}$. Figure~\ref{fig:pixel_track} (b) illustrates a merged cluster produced by two close-by particles that both deposit charge in the second pixel. The fourth pixel is not read-out as the charge is below the read-out threshold. In figure~\ref{fig:pixel_track} (c), a large cluster created from single charged particle is shown. In this case, the cluster is large due to the emission of a $\delta$-ray. $\delta$-rays can also change the amount of collected charge without changing the size of the cluster. In the case of the emission of a $\delta$-ray the cluster position is biased with respect to the true intersection point of the track. 

A connected component analysis (CCA)~\cite{bib:CCA} is used to find groups of neighbouring pixels. This is referred to as eight--cell connectivity because two pixels having a one common corner are considered neighbouring and a single pixel is surrounded by, at most, eight cells. The group of connected pixels is referred to as cluster and information from all pixels is used to estimate the position using a linear approximation. 

Measurements are refined during the fitting of the tracks by using the incident angle and predicted position of incidence on the pixel module of the track associated with the cluster. Such calibrations include a correction for module distortions~\cite{Andreazza:1097395} that cause silicon modules to not be perfectly planar as assumed in the detector description used for Monte Carlo simulation and event reconstruction.

Tracks including all components of the ID are designed to provide a transverse impact parameter resolution of $\sigma_{d_0} = 140~[ \mu\mathrm{m~GeV}] /p_\mathrm{T}[\mathrm{GeV}] \oplus 10~\mu $m and a longitudinal impact parameter resolution of $\sigma_{z_0 \sin{\theta}} = 209~[\mu\mathrm{m~GeV}]/p_\mathrm{T}[\mathrm{GeV}] \oplus 91~\mu $m~\cite{Aad:2008zzm}. The symbol $\oplus$ denotes the sum in quadrature of the two components.\footnote{The impact parameter resolutions depends on $\eta$; the parameterisation shown here is an approximate one, averaged over the full acceptance.}

The charge collected in each of the pixels of the cluster is used to refine the estimate of where the particle intersects the module using a charge interpolation technique, which significantly improves the resolution compared to the purely geometrical limit determined by the pixel size. This technique starts from the geometrical centre of the cluster, defined as ($x_\mathrm{centre},y_\mathrm{centre}$) in the local reference frame of the sensor surface. The charge measurements in the first and last rows and columns of the cluster are used to perform the interpolation.

The particle intersection is estimated by using the following equations:
\begin{eqnarray}
x_{\mathrm{cluster}} & = & x_{\mathrm{centre}} + \Delta_\mathrm{x} (\phi, N_{\mathrm{row}}) \cdot \left[ \Omega_\mathrm{x} - \frac{1}{2} \right] \\
y_{\mathrm{cluster}} & = & y_{\mathrm{centre}} + \Delta_\mathrm{y} (\theta, N_{\mathrm{col}}) \cdot \left[ \Omega_\mathrm{y} - \frac{1}{2} \right],
\end{eqnarray}
where~$\Omega_\mathrm{x(y)}$ is defined as
\begin{eqnarray*}
\Omega_\mathrm{x(y)} & = & \frac{q_{\mathrm{last\ row (col)}}}{q_{\mathrm{first\ row (col)}}+q_{\mathrm{last\ row (col)}}},
\end{eqnarray*}
and $q$ denotes the charge collected in a given pixel. The parameter $\Delta_{\mathrm{x}}$ ($\Delta_{\mathrm{y}}$) is a function of the projected incident angle with respect to the sensor surface, $\phi$ ($\theta$), and the number of pixels within the clusters in the $x$ ($y$) direction, $N_{\mathrm{row}}$ ($N_{\mathrm{col}}$). These parameters are extracted from either Monte Carlo simulations or data~\cite{Aad:2010bx}. The charge interpolation technique was the default ATLAS clustering approach until mid-2011 and will be referred to as CCA clustering.
 
\begin{figure}[htb!]
\begin{center}
\includegraphics[width=0.8\textwidth]{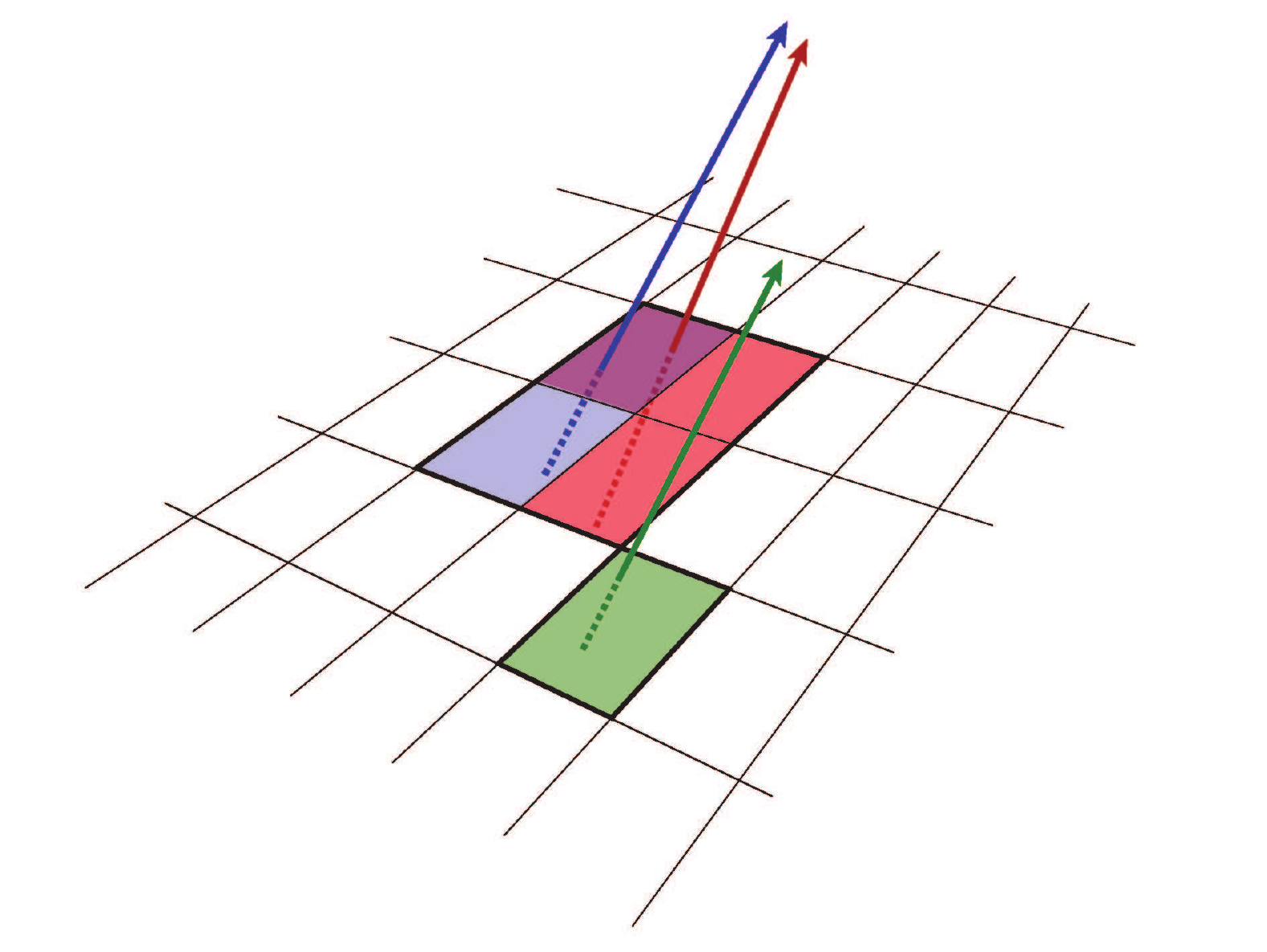}
\caption{Illustration of charge deposited by multiple particles in the dense core of a jet in a layer of the pixel detector. The pixel size is not drawn to scale. The arrows indicate the passage of charged particles through the pixel sensor. The pixels are shaded according to which particle deposited charge in them. The dashed lines indicate the path traversed by the particles in the silicon and the solid line shows the single cluster obtained by the eight-cell CCA.\label{fig:track_density}\label{fig:fig_group_problem}}
\end{center}
\end{figure}

In dense environments, such as in the cores of energetic jets, there is an increased risk that pixels traversed by different particles are merged by the CCA, or even that charge induced by several particles is collected in a single pixel. This problem occurs more and more often as the spatial separation between the particles on the sensor plane approaches the pixel size. This cluster merging is depicted in figure~\ref{fig:track_density}, which illustrates an event in which the charge induced by three particles is reconstructed as a single cluster.

\begin{figure}[htb!]
\begin{center}
\includegraphics[width=0.8\textwidth]{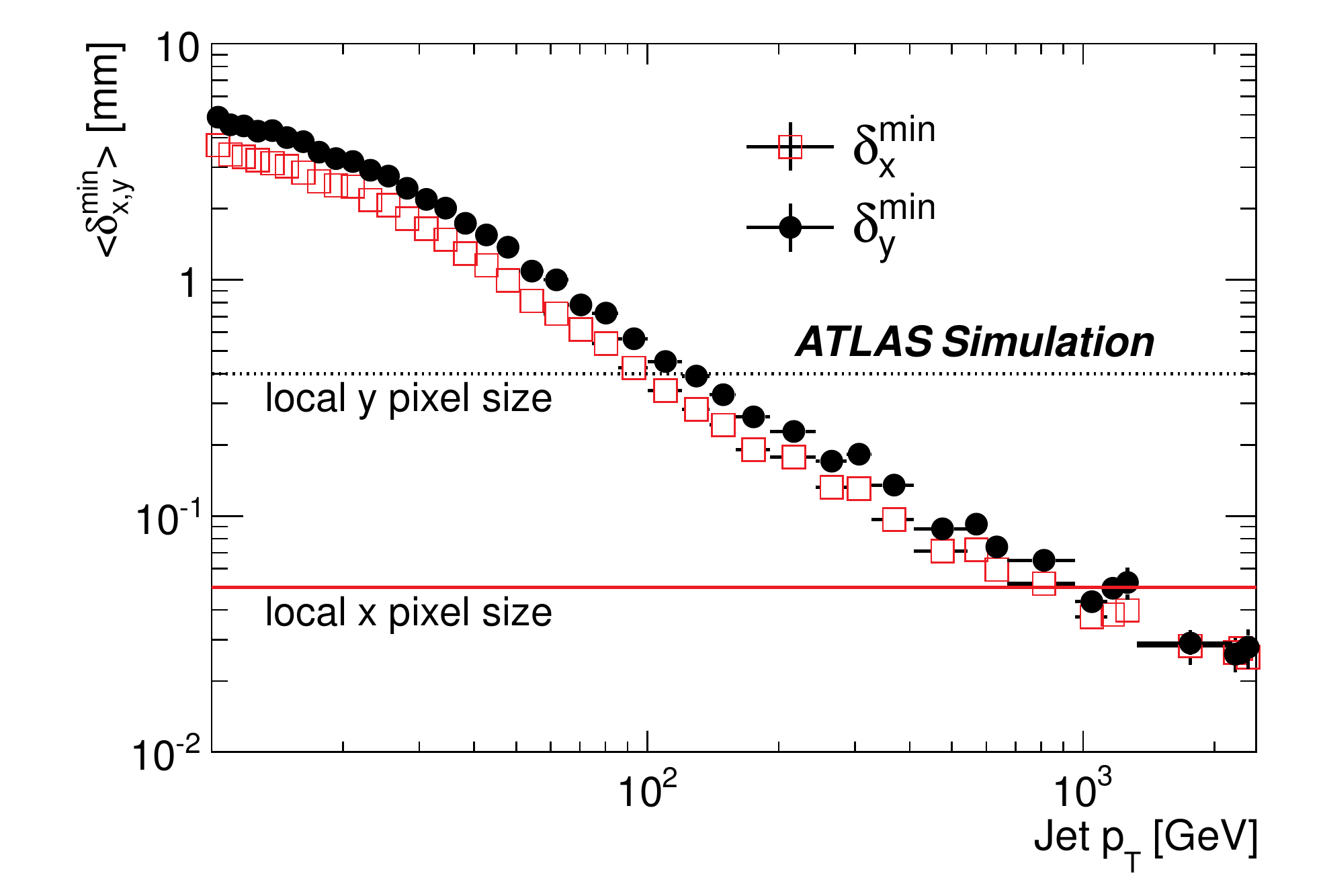}
\caption{Average separation between the two closest charged particles in a jet in the transverse (\protect $<\delta_\mathrm{x}^{\mathrm{min}}>$, open squares) and longitudinal (\protect $<\delta_\mathrm{y}^{\mathrm{min}}>$, full circles) direction at the innermost layer of the pixel barrel. This is shown as a function of the transverse momentum of the jet. The pixel size in the transverse (50 $\mu$m) and longitudinal (400 $\mu$m) direction is indicated with the solid and dotted lines, respectively. \label{fig:dRMin}}
\end{center}
\end{figure}

Figure~\ref{fig:dRMin} shows the average separation in the transverse ($\langle\delta_\mathrm{x}^{\mathrm{min}}\rangle$) and longitudinal ($\langle\delta_\mathrm{y}^{\mathrm{min}}\rangle$) direction of the two closest stable charged particles in jets, at the radius of the innermost pixel layer in the barrel. Only track pairs separated by less than a pixel in the longitudinal (transverse) direction are shown here for the transverse (longitudinal) direction. A sample of simulated dijet events based on the {\sc PYTHIA}~\cite{pythia} Monte Carlo generator with the leading jet $p_\mathrm{T}$ greater than 800 GeV was used. Jets were reconstructed from stable generator-level particles using an anti--k$_t$ jet algorithm~\cite{AntiKt} with a cone size of 0.4. The figure illustrates that shared measurements appear already in jets with relatively moderate momentum as cluster merging starts before the pixel size is reached. In the worst case, when cluster merging appears in pixel layers beyond the innermost and the number of shared measurements on a track exceeds the given threshold, the track candidate is completely disregarded to avoid the creation of duplicate tracks. This leads to an inefficiency in finding both tracks. The limit where two close--by tracks can still be reconstructed separately is often referred to as double--track resolution. With the CCA clustering, no attempt is made to identify or split these merged clusters.
%At a later stage of the track reconstruction, merged clusters are identified and flagged as such through their appearance on multiple track candidates.  

% The intrinsic geometrical limit that allows to separately reconstruct close-by tracks is often referred to as double track resolution. 
%The Inner Detector~\cite{Aad:2010bx} comprises three complementary sub-detectors: the Pixel Detector, the Semi-Conductor Tracker (SCT) and the Transition Radiation Tracker (TRT). The TRT uses straw tubes which are capable of particle identification using transition radiation, while the SCT uses silicon strips and the pixel detector silicon pixels.
%The ATLAS pixel detector has sensitive elements covering radial distances between 50.5~mm and 150 ~mm. The modules are arranged into three concentric barrel layers and two endcaps each of three disks to provide typically three measurement points per track. Most of the $\sim$80.4 million individual pixels are 50~$\mu$m $\times$ 400~$\mu$m. The shorter pitch is arranged to provide a high-precision measurement in the radial plane. Hits in a pixel are read out if the signal exceeds a tunable threshold. The pixel detector provides a charge measurement through measuring the pulse height using the time-over-threshold (ToT) technique.

\section{Pixel cluster splitting}
\label{sec:pix_clus_split}
If merged clusters from several charged particles are split into sub--clusters for each particle they can appear as individual measurements on tracks. This improves the double--track resolution and reduces the number of measurements shared between tracks. The identification of such merged clusters without performing any splitting can already improve track quality because this allows for a dedicated treatment in the measurement calibration or ambiguity-resolving process (see section~\ref{sec:track_reco}). 

Both the cluster size and the charge collection pattern of the cluster can be exploited when attempting to split the cluster. Assumptions about the particle origin and direction are required to determine the predicted cluster shape. %Previously, a cut based cluster splitting approach based on the charge collection pattern in the transverse direction was deployed for the track reconstruction of the CDF experiment~\cite{CDFTrackReconstruction}.
 
\subsection{A neural network for cluster splitting}
\label{sec:NNclusterReco}

Artificial neural networks are powerful tools for solving complex pattern recognition problems characterised by significant non--linearities. The increasing CPU power available for event reconstruction in high--energy physics makes them attractive for problems with many degrees of freedom. A novel approach to clustering based on artificial neural networks is presented. A single NN is used to estimate the probability that a cluster was created by one or many particles and to split the cluster when possible. Two sets of NNs are used to estimate cluster positions and uncertainties, containing three and six NNs respectively. This approach allows the NN clustering algorithm to also improve the cluster position estimation and hence the resolution of the track parameters.

Clusters are first created by running the CCA algorithm. The NN clustering algorithm then runs over all the resulting clusters and obtains an estimate of:
\begin{enumerate}
\item the number $N$ of charged particles traversing the cluster;
\item the two--dimensional $\vec P_{i}$ (with $i={1,N}$) of the positions and corresponding uncertainties of the impact points of the particles on the pixel sensor.
\end{enumerate}

Only clusters which are contained in a rectangular matrix of 7$\times$7 pixels are included. If any pixel is further than three pixels from the centroid, the cluster is not processed by the NN clustering algorithm. There are only less than 1\% of such clusters and they most probably are due to particles not originating from the beam collisions. In data there is an increased rate of large clusters from beam backgrounds and low--momentum particles. The latter is because these are below the threshold in the simulation. Neglecting big clusters in the NN clustering algorithm does not influence the high \pT\ tracking performance as these clusters originate from the charged particles of \pT<0.4 \GeV\ which are below the track reconstruction threshold. No attempt is made to assign single pixels of the original cluster to the split cluster. This is because the pixel content of a cluster itself has no influence on the track reconstruction after it has been exploited to determine the cluster position and error.

In the first pass of the neural network clustering algorithm, particles are assumed to have been produced at the centre of the beam interaction region, which is estimated using the primary vertices of collisions from a sub--sample of events analysed during a rapid first pass of the reconstruction algorithms. The incident angle of the particles is calculated using a straight line from the centre of the beam interaction region to the cluster position. The uncertainty on the incident angle in the longitudinal direction can be large. Clusters in pixel modules at small radii and covering a region of small $|\eta|$ are affected more than clusters in modules at high radii or at high $|\eta|$. This is exploited by the neural network in order to reduce such uncertainty. Once track candidates have been identified, the algorithm is rerun to benefit from the significantly refined estimate of the incident angle. Different neural networks are trained and used in the two cases. No additional attempt to split clusters is made at this stage, because the initial pattern recognition stage of the track reconstruction is only run once, which means that this additional information cannot be used to identify further track candidates.

A multi--layer perceptron, which is a feed--forward artificial neural network, is used. This is equivalent to a set of functions with input variables (nodes) $x_k$ ($k=\left[0,N_{\mathrm{inputs}}\right]$) and output values (nodes) $F_i$. It has two intermediate hidden layers and can be denoted by:

\begin{eqnarray}
F_{i}\left(\vec x\right) & = & h\bigg( \sum_j \omega^2_{ij} g\Big( \sum_{k} \omega^1_{jk} g\big(\sum_{l} \omega^0_{kl} x_{l} + \chi^0_{l} \big)+ \chi^1_{j} \Big) + \chi^2_{i} \bigg),
\end{eqnarray}
where: $\omega^{\left(0,1,2\right)}$ and $\chi^{\left(0,1,2\right)}$ correspond to the weight and threshold parameters which are adjusted during the training process, with the upper index denoting the layer number; $g(x)$ and $h(x)$ are the activation functions for computing the values of the nodes in the intermediate and output layers. Here $g(x)$ is chosen to be:
\begin{eqnarray} 
g\left(x\right) & = & \frac{1}{ 1 + \mathrm{e}^{-2 x} } .
\end{eqnarray}

The choice of $h(x)$ depends on the network. For the estimation of the number of particles traversing a cluster, $h(x)$ takes the same form as $g(x)$, which confines the output nodes to be between 0 and 1, while for the estimation of the cluster position, linear functions are chosen ($h(x)=x$). All networks are trained using the JetNet package~\cite{Peterson:259970}. This package minimises the sum of the mean squares of the errors of the network output nodes with respect to their target values. The sum is taken over all patterns in the training set. 

%This is possible because the ID track reconstruction strategy allows measurements to be recalibrated after they have been assigned to a track candidate.

\paragraph{Number of particles per cluster}
The number, $N$, of particles in a cluster is estimated using a neural network dedicated to classification.
It is assumed that a cluster may be split into parts stemming from at most three particles, thus the number of output nodes is chosen to be equal to three and the target is chosen pattern by pattern to be $\left\{ 1 0 0 \right\}$, $\left\{ 0 1 0 \right\}$ or $\left\{ 0 0 1 \right\}$ for the cases of one, two or three traversing particles, respectively. This ensures that the output nodes $F_{i}$ can be interpreted as the posterior Bayesian probabilities for a pattern with inputs $\vec x$ to be of the type $i$, provided the training converged to the global minimum. A higher number of particles per cluster is not considered by the NN clustering algorithm. The Bayesian probabilistic interpretation relies on both the description of observed cluster properties and the composition of the training sample. The posterior probability of having either two or three particles traversing the cluster is referred to as $P_\mathrm{shared}$. 

In the simplest approach, the candidate cluster properties would be condensed into a few input variables to the NN, which would then be trained to distinguish between the different output types. Instead, a more aggressive approach is followed here by feeding the neural network input nodes $x_k$ with the full available information:
\begin{itemize}
\item a fixed--size 7$\times$7 matrix of the charge collected in each pixel of the candidate cluster,
\item a fixed--size vector of seven elements with the longitudinal size of the pixels in the 7$\times$7 matrix to label the long pixels,
\item layer number (from 0 to 2) and layer type (barrel or endcap),
\item the angles of incidence $\phi$ and $\theta$ of the candidate charged particle with respect to the sensor surface,
\item $\eta$ of the pixel module (only if no track candidate is available yet): 
\end{itemize}

In total, this corresponds to 60 (or 61) input nodes. If a pixel in the charge matrix has charge below the readout threshold, the matrix entry is set to zero. The matrix is centred at the position of the pixel containing the charge--weighted centroid of the cluster.

Clusters with $P_\mathrm{shared} > 0.5$ are split, while clusters with $P_\mathrm{shared} > 0.02$ are labelled as merged clusters. The optimisation of the cut to split clusters is discussed in section~\ref{sec:nnperf}.

\paragraph{Cluster position and error estimate}

The cluster position is estimated using a second set of neural networks configured for interpolation. In this case, neural networks are used to obtain a function for which the outputs are as close as possible to one or more continuous target values by exploiting the dependence of such targets on the input variables. Three different neural networks are used depending on the number of particles $N$ (from 1 to 3) traversing the candidate cluster and the corresponding training sets are selected based on the true number of particles in the simulation. All networks use the same input variables as the classification network, while the $2 \times N$ output nodes are trained to represent an estimate of the two--dimensional position of the $N$ particle(s) impact point(s) $\vec P_\mathrm{i}$. For each pattern, the targets in the training are set to the desired continuous values. 

A third set of neural networks is used to estimate the probability density function (pdf) describing the residual of the estimated point of impact, $\Delta \vec P = \vec P - \vec P_\mathrm{true}$. This is the difference between the reconstructed, $\vec P$, and true, $\vec P_\mathrm{true}$ points of impact.

This pdf therefore depends on all cluster properties encoded in the input nodes. The impact point position(s) relative to the cluster centroid are used as additional inputs for training. Different neural networks are used to estimate the uncertainties in the transverse (longitudinal), $NN_\mathrm{err_x}$ ($NN_\mathrm{err_y}$) directions. Different neural networks are trained depending on the number of traversing particles $N$ (from 1 to 3). This leads to six neural networks in total. The possible continuous outcomes of $\Delta P_\mathrm{x}$ ($\Delta P_\mathrm{y}$) are subdivided into intervals and one output node is mapped to each interval, reducing the problem of estimating the probability density function to a classification problem. 
If a training pattern obtains a deviation $\Delta P$ corresponding to interval $j$ in the training, then the target values of the output nodes will be $ \left\{ 0 ... 0 1 0 ... 0 \right\} $, with the value of one filling the $j$--th position.
 
This is extended to multiple particles by considering $N$ sets of intervals, each set corresponding to one of the $N$ particles, resulting in $N \times N_\mathrm{bins}$ output nodes. As a consequence, for each pattern, $N$ of the output nodes are set to one. Therefore the output nodes yield the posterior Bayesian probability that, given the observed cluster, the estimated position results in a deviation from the true position within the interval represented by that specific bin.\footnote{The maximum range in $\mu$m considered for $\Delta P_\mathrm{x}$ ($\Delta P_\mathrm{y}$) in $\mu$m is $[-30,30]$ ($[-300,300]$) for the one particle case and $[-50,50]$ ($[-400,400]$) for the remaining cases with two or three particles. The granularity is defined by $N_\mathrm{bins}$, which is chosen to be 30, 25 and 20 for the 1, 2 and 3 particle cases, respectively.}

%The probability density function for $\Delta \vec P$ is a smooth function of the estimated position $\vec P$. 
This set of neural networks yields a cluster--by--cluster estimate of the binned pdf for the deviation of the estimated position(s) from the true one(s). In the track reconstruction, cluster measurements are assumed to follow a Gaussian distribution and therefore only a single value for the position uncertainty is needed. This is derived for each traversing particle as the square root of the second moment of the distribution. In the future, for specific applications such as $b$--tagging or $\tau$ reconstruction, the full error pdf could be used as input to more complex track fitting approaches such as the Gaussian Sum Filter~\cite{Fruhwirth:1996be}. Such approaches allow the assumption of Gaussian--distributed measurements to be dropped.

\subsection{Neural network training}
\label{sec:NN_Training}

Ten neural networks are needed for up to $N=3$ possible particles: $NN_\mathrm{par}$ estimates the number of particles traversing
a cluster and the remaining nine estimate the positions ($NN_\mathrm{pos}$) and errors ($NN_\mathrm{err_x}$ and $NN_\mathrm{err_y}$). The training data are a mixed sample 
of top--antitop pair production ($t\bar{t}$) and highly energetic dijet events ($140<p_\mathrm{T}^\mathrm{jet}<560$~GeV). The $t\bar{t}$ events were generated using the
{\sc MC@NLO}~\cite{Frixione:2002ik} generator interfaced with {\sc HERWIG}~\cite{herwig} to model the parton shower and the dijet events were generated using {\sc PYTHIA}. Both samples
were passed through the standard {\sc Geant4}~\cite{Agostinelli:2002hh} simulation of the ATLAS detector~\cite{Aad:2010ah} and the pixel digitisation model. The detector simulation includes effects such as 
ionisation loss, the production of knock--out shell electrons, $\delta$--rays, and hadronic spallation from the interactions of charged particles in the pixel modules. The digitisation model accounts for the charge drift to the readout surface including the Lorentz angle and effects from thermal diffusion, cross--talk between neighbouring channels and the modelling of the ToT shape and its 8--bit readout precision. Time--dependent radiation effects are not included at this stage, but may need to be considered in the future when the detector will have suffered from significant radiation damage.

All networks are trained on a number of training patterns which exceed the number of network parameters by at least a factor of a thousand, in order to avoid problems due to the finite size of the training sample. All network layouts include two hidden layers. The inputs and continuous outputs are shifted and normalised to yield an average value of zero and a standard deviation of one to improve the efficiency of the neural network training. The data used to train the neural network are subdivided into a training and a testing sample. The training of each network is completed once the sum of the mean square errors of the target values with respect to the values estimated by the network reaches a minimum, when computed on the statistically independent testing sample. A typical training takes from 12 to 24 hours on a single CPU.

\subsection{Changes to the track reconstruction}
\label{sec:track_reco}
The NN clustering algorithm allows many previously merged clusters from two nearby particles to be resolved. Consequently, the track--finding strategy and the candidate selection cuts in the ambiguity processing is modified to take advantage of the improved double--track resolution.

Prior to the NN clustering, track candidates were found using a progressive filter that iteratively selected the best matching measurement on each layer, starting from an initial track seed. With the NN clustering, the progressive filter is not optimal as it does not consider all possible nearby clusters in a pixel module, therefore a combinatorial filter is used instead. The combinatorial filter considers more than one measurement on a predicted measurement layer when building the track candidate, which may lead to several track candidates per track seed. The extrapolation of low momentum tracks has large uncertainties due to multiple scattering. Therefore, all split clusters from the original cluster found by the CCA are considered in the combinatorial filter if their separation is within the extrapolation uncertainty. In order to reduce the CPU overhead, seeds that are already contained in previously reconstructed candidates are not used to seed new candidates.

The track selection in the subsequent ambiguity processing also benefits from the significantly reduced rate of shared measurements to efficiently reject fake candidates. After running the NN clustering algorithm, a track can share a measurement with another track only if the cluster is not split and the output of the NN is compatible with a possible merged cluster. 

\section{Performance of the neural network clustering algorithm}
\label{sec:nnperf}

\begin{figure}[htbp!]
\begin{center}
\includegraphics[width=0.9\textwidth]{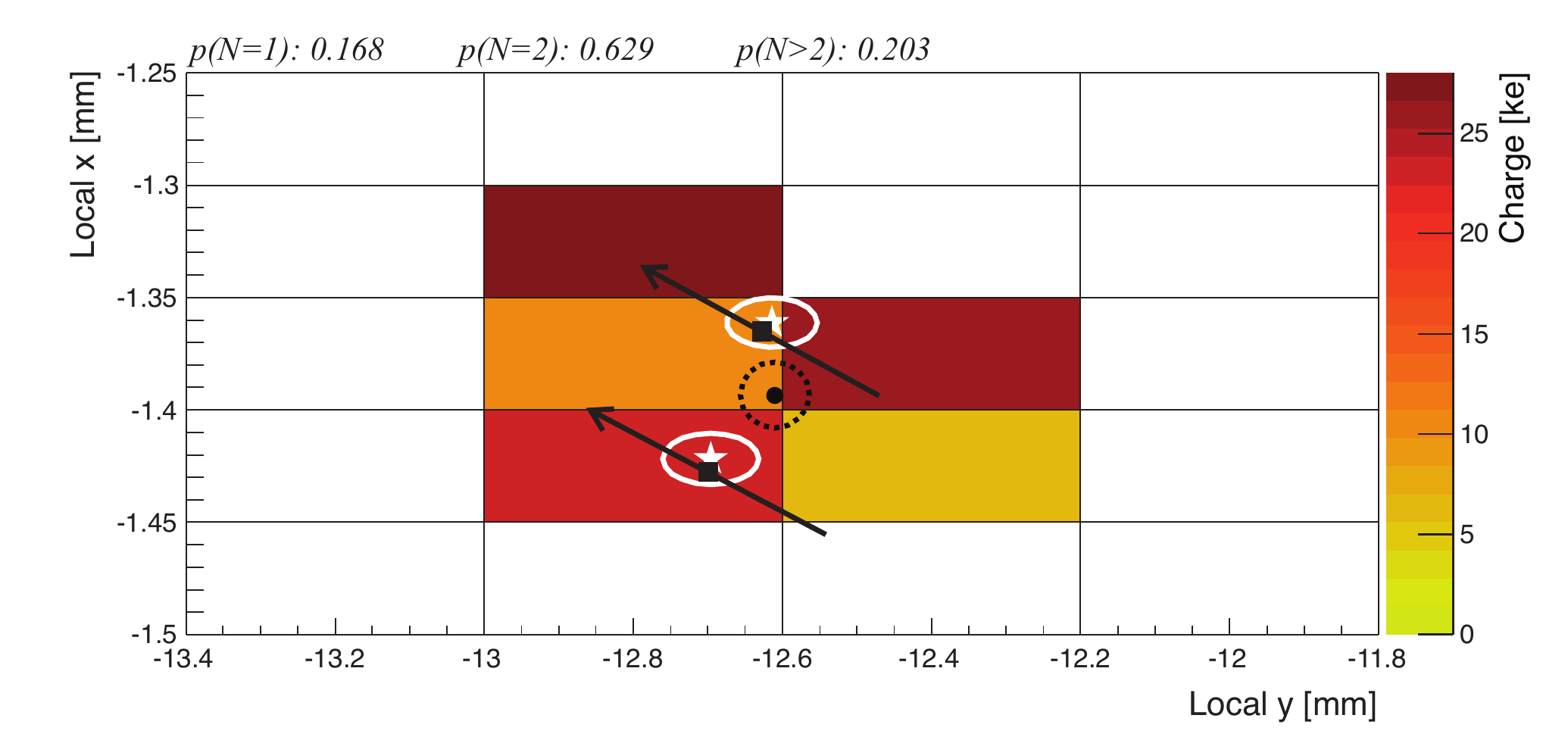}
\caption{Example of a merged cluster created by two particles, from Monte Carlo simulation. The two arrows show the paths of the particles through the silicon, the black squares indicate the true intersection with the mid--sensor plane. The black dot illustrates the non--split cluster position, while the two stars show the estimated cluster positions after splitting. The ovals indicate the position error estimates and the $p(N=i)$ denote the probabilities for the cluster to have been created by $i$ particles as estimated by the neural network. Effects caused by the Lorentz angle in the silicon sensor were removed in this illustration.\protect\label{fig:exampleSplit} 
}
\end{center}
\end{figure}

Figure~\ref{fig:exampleSplit} illustrates the result of a successful split achieved by the NN clustering algorithm on an example cluster created by two close--by particles from Monte Carlo simulation. In this figure, the black arrows indicate the paths taken by the true particles passing through the silicon. The black circle surrounded by a dotted ellipse indicates the single cluster position and error estimate as obtained by the CCA clustering. The white stars show the cluster positions obtained from the NN clustering algorithm and the white ovals indicate the error estimates. The clusters identified by the NN clustering reproduce very well the paths of the true particles.

\begin{figure}[htb!]
\begin{center}
\includegraphics[width=0.8\textwidth]{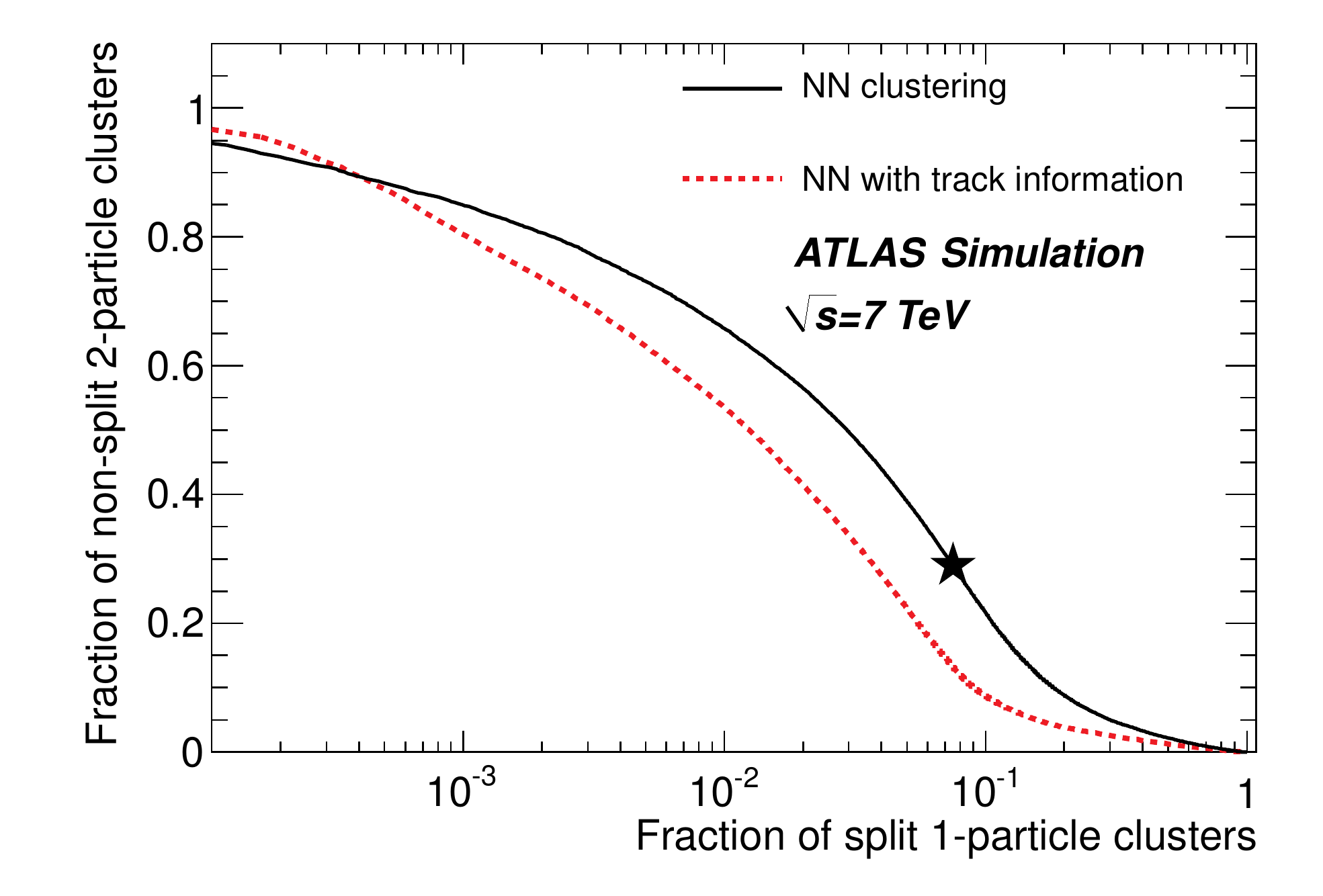}
\caption{The fraction of non--split two--particle clusters versus the fraction of incorrectly split one--particle clusters in simulation. The cut value of the NN varies along the solid and dashed lines: a tight cut corresponds to the lower right corner of the plot and a loose cut correponds to the upper left corner. The distribution is shown both for the NN clustering algorithm using only the cluster and luminous region information (solid line) and for the NN clustering algorithm additionally including the track information (dashed line). The chosen working point is indicated with a star for the setup without track information, because, due to the configuration of track reconstruction algorithms used by ATLAS, this is the only pass of the NN algorithm used to split clusters.\protect\label{fig:splitperfor} 
}
\end{center}
\end{figure}

There is, however, the possibility that the algorithm falsely splits clusters. False splitting raises the risk of creating fake or duplicate tracks. The relative fractions of correctly split to falsely split clusters was studied using Monte Carlo simulations, where the number of contributing particles to each cluster is known. Figure~\ref{fig:splitperfor} compares the interdependence between falsely split clusters stemming from a single particle and the rate of clusters created by two particles which were not split. The chosen working point of the NN clustering splits about 71\% of the clusters that arise from 2 particles correctly. On the other hand, 7.5\% of the clusters that arise only from one particle are incorrectly split into two, which is indicated in the figure by a star.

The cluster--splitting performance can be significantly improved when the incident angles from reconstructed tracks are used instead of estimates based on the centre of the luminous region. This is also seen in figure~\ref{fig:splitperfor}, which shows that using track information improves the rate of incorrectly split clusters by approximately a factor of two at 71\% efficiency. However, in the current ID track reconstruction strategy, no additional track finding is run after the track fitting stage and therefore an additional pass of the splitting algorithm cannot be used to resolve shared clusters or recover previously lost tracks. Therefore, the $NN_\mathrm{par}$ run with the track information are only used to improve the resolution and error estimates.

\begin{figure}[htb!]
\begin{center}
\includegraphics[width=0.49\textwidth]{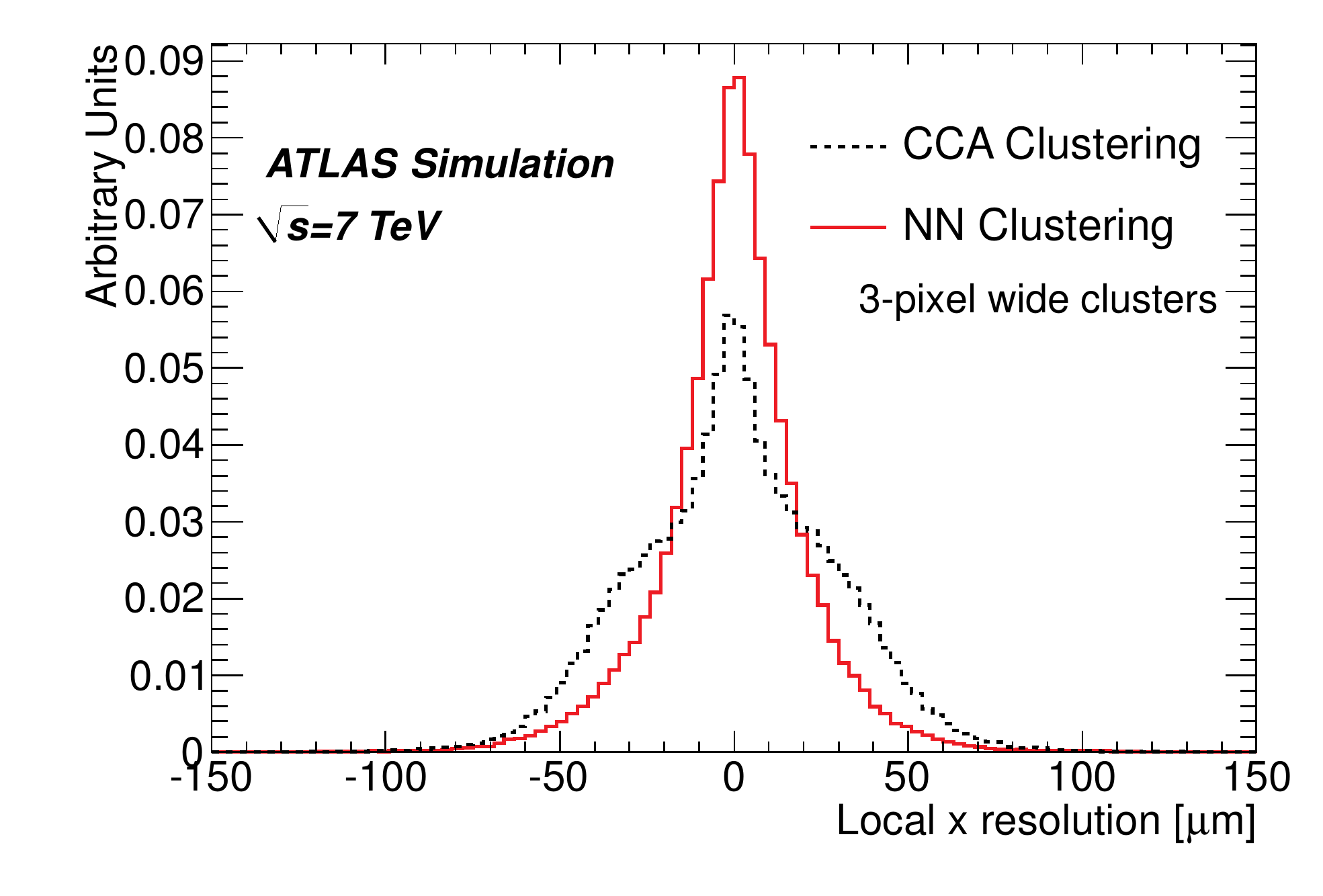}
\includegraphics[width=0.49\textwidth]{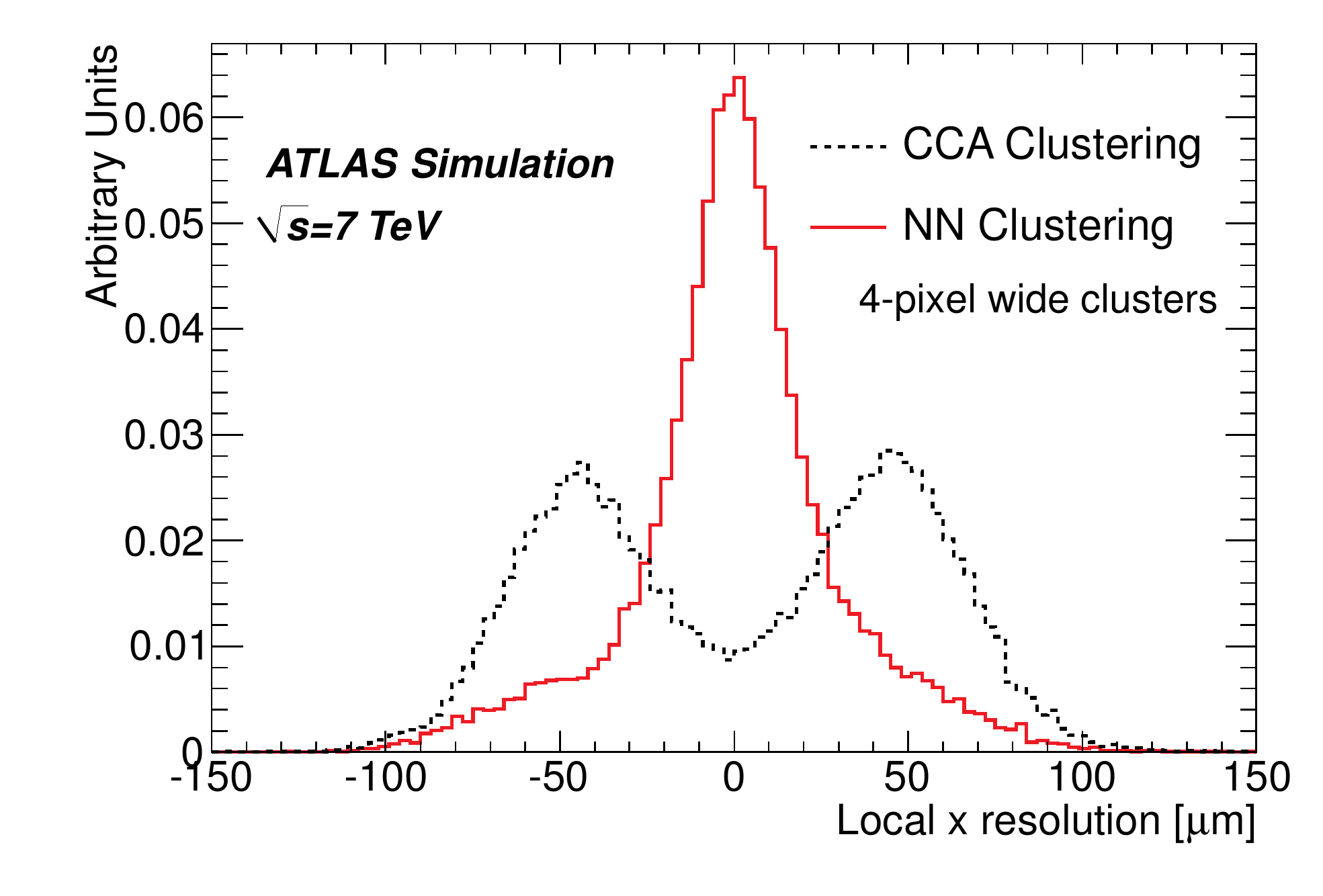}
\caption{The cluster residual in the local $x$ direction for clusters with a width of three (left) or four pixels (right) in the $x$--direction reconstructed with the CCA clustering algorithm (dashed line) and the NN clustering algorithm (solid line).}\label{fig:PhiResiduals}
\end{center}
\end{figure}

Merged clusters, when assigned to a track, usually yield a degraded position estimate. This results in a larger spread of the track--to--cluster distance, the so--called cluster residual. Figure \ref{fig:PhiResiduals} shows the comparison of the residuals in the local $x$ direction for clusters with a width of three or four pixels in the transverse direction when applying the NN clustering algorithm, compared to the outcome of the CCA algorithm. A clear improvement can be seen in both cases and becomes particularly visible in the four--pixel wide cluster category. The layout of the ATLAS pixel detector minimises the probability that a reconstructable charged particle originating from the luminous region traverses more than three pixels in the transverse direction as discussed in section~\ref{sec:cluster}. Thus, most of those clusters stem from multiple particles or $\delta$--rays. Both cases lead to a double peak structure reflecting the 50 $\mu$m pitch size in the transverse direction. This double peak structure completely vanishes with the NN clustering algorithm. Some non--Gaussian tails are present in the residuals for the NN clustering.  These originate from large--angle scattering, $\delta$--rays (which are not included in the labelling of the number of particles per cluster in the NN training sample) and from clusters on the edges of modules, which typically have skewed distributions.

\subsection{Performance in data and simulation}

The MC simulation was used to obtain the pattern of charge deposition for the training set. Good performance of the NN clustering algorithm, as well as a good agreement between the performance on data and simulation, depends on how well the interaction of particles with the silicon and the signal collection is modelled by the detector simulation and digitisation. Cluster merging depends on the local charged--particle density, Lorentz drift and the incident angles of the traversing particles; effects from charge collection and channel cross-talk are negligible. Clusters in the barrel and endcap are thus treated similarly, but with the detector region given as input to the NN, so cluster classification is performed based on cluster sizes.

\begin{figure}[htb!]
\begin{center}
\includegraphics[width=0.49\textwidth]{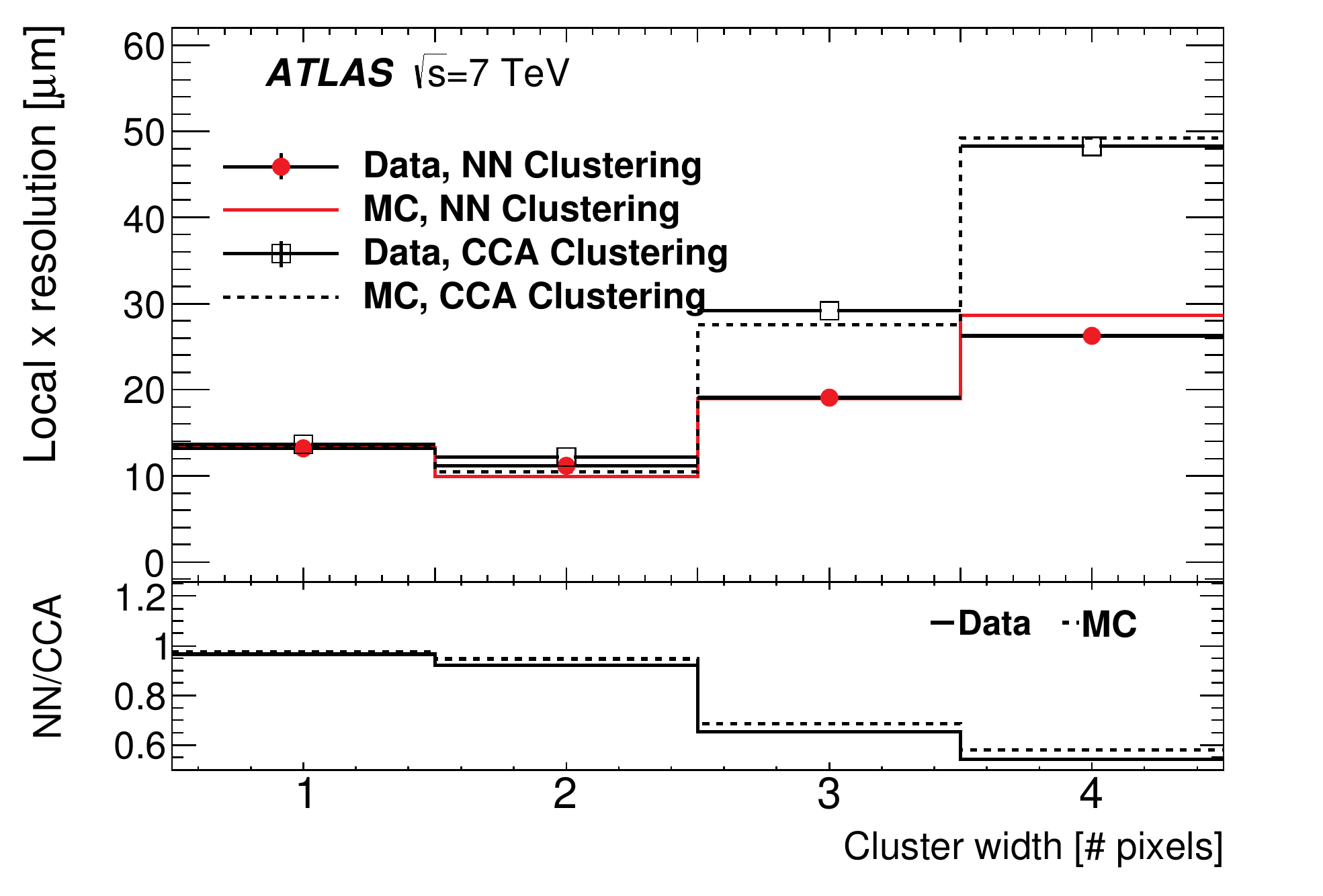}
\includegraphics[width=0.49\textwidth]{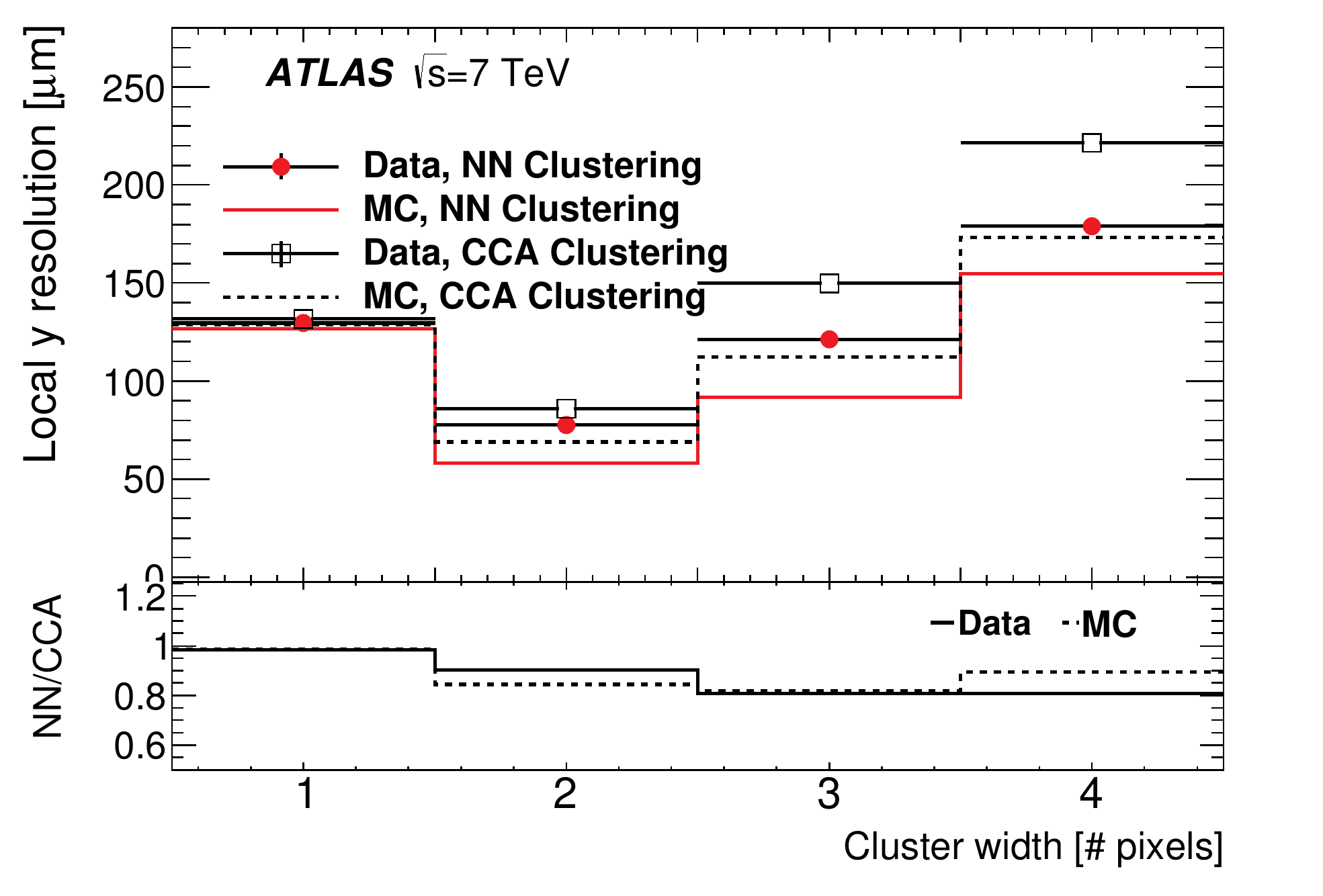}
\caption{Measurement residual resolution in the transverse (left) and longitudinal (right) direction in data and simulation for the CCA and NN clustering algorithms. The lower panel shows the ratios of the residual resolutions of the NN compared to the CCA clustering algorithms. Error bars are included for the data but are smaller than the markers.\label{fig:hit_residuals}}
\end{center}
\end{figure}

Figure~\ref{fig:hit_residuals} compares the root mean square (RMS) of the measurement residuals for the CCA clustering and the NN clustering algorithm in data and simulation in the transverse and longitudinal direction in the different cluster categories. The majority of three-- and four--pixel wide clusters in the transverse direction are due to close--by particles and $\delta$--rays. In the longitudinal direction, clusters of this size are geometrically possible due to the shallower incidence angle. The improvement shown in figure~\ref{fig:hit_residuals}(left) can thus be mostly attributed to actual cluster splitting, which includes splitting components from $\delta$-rays, while in figure~\ref{fig:hit_residuals}(right) a sizeable contribution of the improvement is caused by the non--linear charge interpolation of the NN clustering algorithm. Discrepancies between data and Monte Carlo simulation can arise from imperfections of the detector such as module misalignment or deformations that are not present in the simulated model of the detector geometry,
as well as from limitations in the detector simulation and digitisation model that include several complex components as described in section~\ref{sec:NN_Training}. Discrepancies are seen in figure~\ref{fig:hit_residuals} for the longitudinal direction. This is most likely due to limitations in the modelling of the longitudinal charge sharing. Nonetheless, the relative improvement obtained by the NN clustering algorithm compared to the CCA clustering algorithm is largely consistent between data and Monte Carlo simulations.

%It has been shown that large clusters, which have an enhanced probability of containing charge from multiple particles, show the most significant improvement when applying the NN clustering algorithm. However, clusters produced by single particles show a clear improvement in the resolution of the hit residuals is obtained, due to the improved estimate of the cluster position from the neural network. The NN clustering algorithm also brings an improvement of the track resolution for isolated tracks that are not affected by nearby tracks as it is the case in dense environments. This is due to the fact that the NN approach is optimised to interpret the cluster position using the charge deposition matrix of the pixel cluster together with an assumption of the track origin. Also the identification of effects caused by knock-on shell electrons in the silicon, often referred to as $\delta$ rays is possible in the NN clustering, since it is very similar to multiple particles passing close-by through the layer, though this has not been exploited yet.

\begin{figure}[htb!]
\begin{center}
\includegraphics[width=0.49\textwidth]{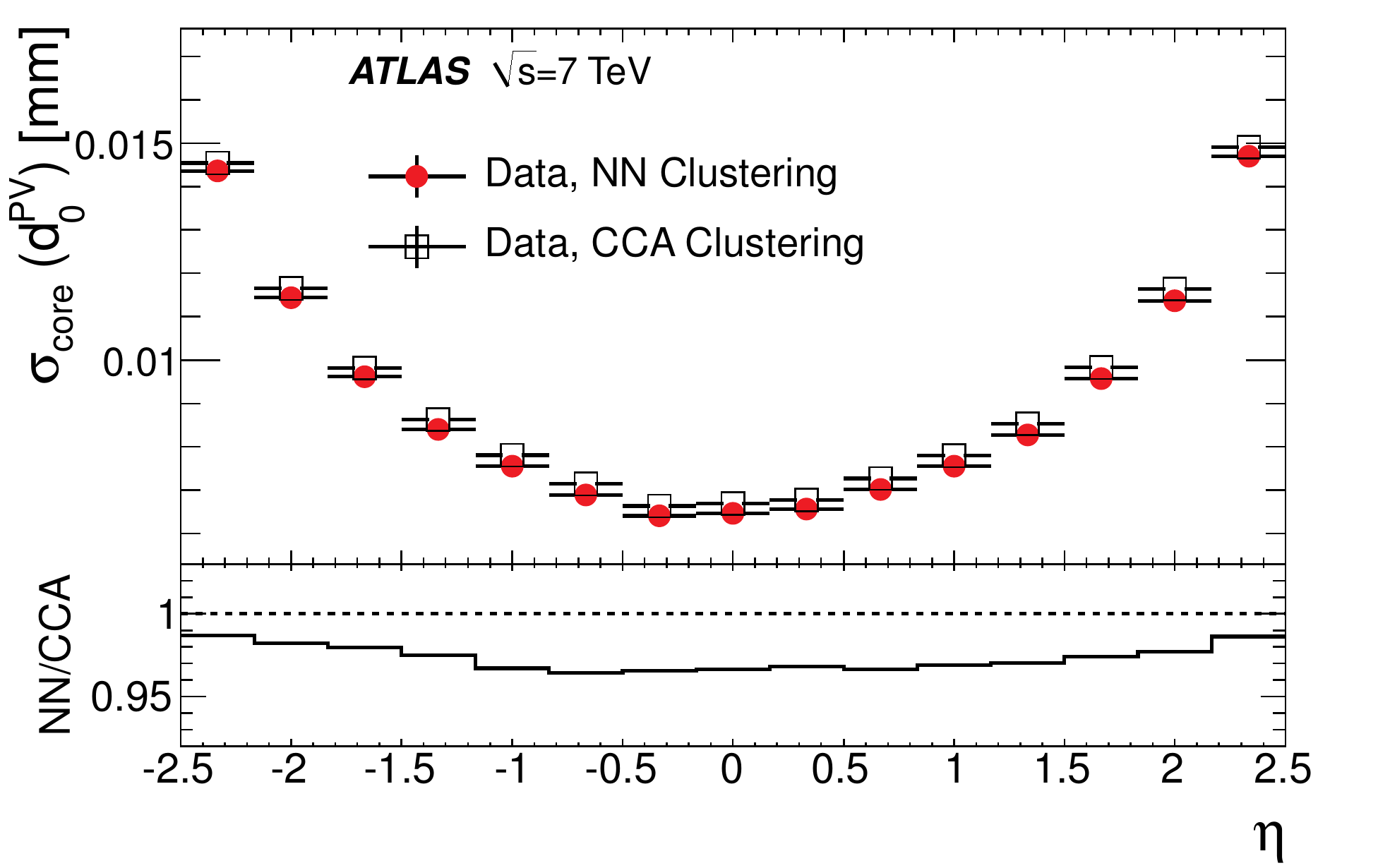}
\includegraphics[width=0.49\textwidth]{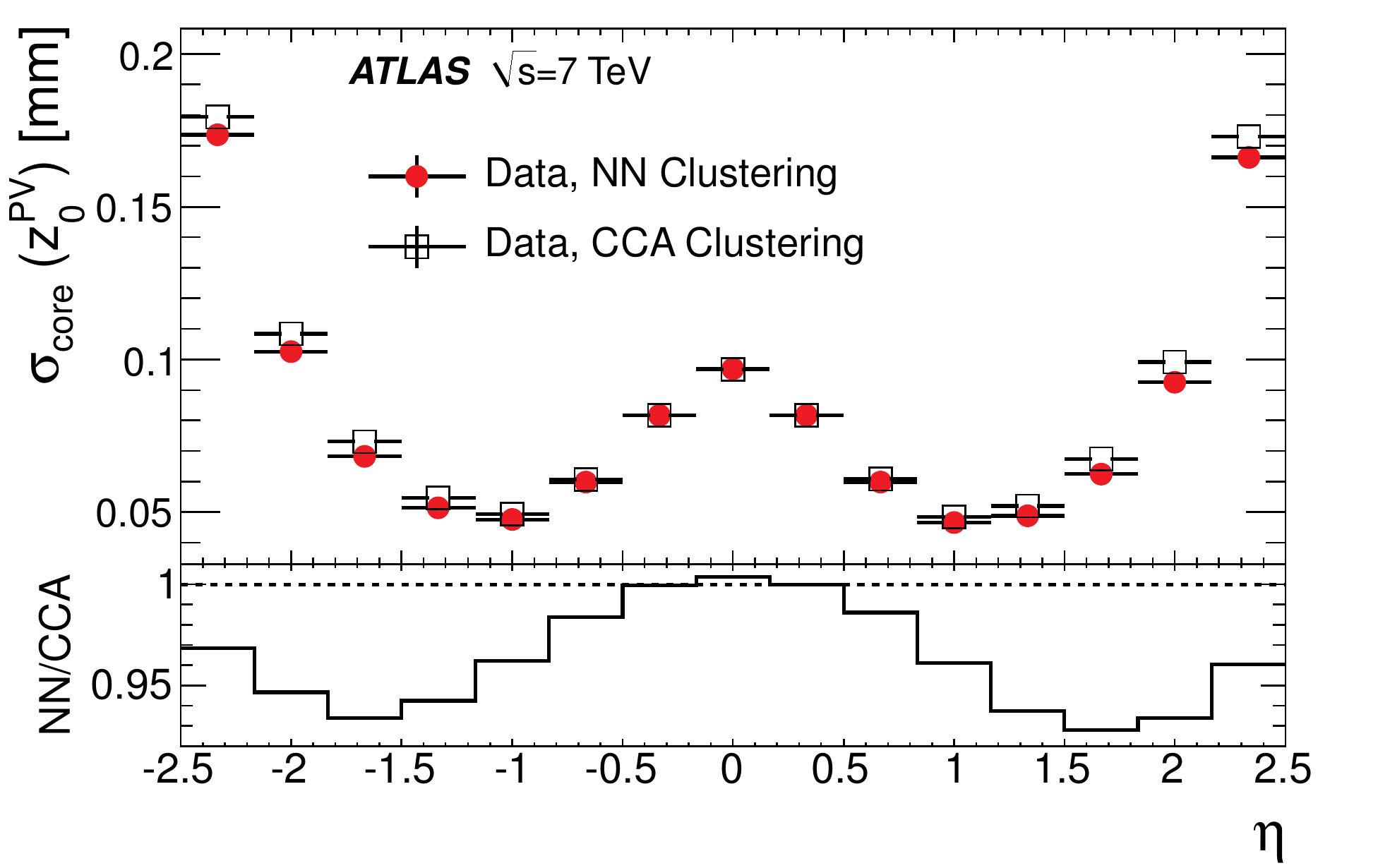}
\caption{The transverse (left) and longitudinal (right) impact parameter resolution with respect to the primary vertex for isolated muons from $Z$ boson decays with the NN (solid) and the CCA (open) clustering algorithms. The ratio of the resolutions obtained with the NN and CCA clustering algorithms is indicated in the lower panels. Error bars are included for the data but are smaller than the markers.
\label{fig:zmumu} }
\end{center}
\end{figure}

The improvement coming from the non--linear charge interpolation and $\delta$--ray handling in the NN clustering can be checked on isolated tracks as there are no other close--by particles from the beam collision. Pairs of oppositely charged combined muons with $p_\mathrm{T} > 25$~GeV, which produce a $Z$ boson candidate with a mass $m_{\mu\mu} > 50$~GeV were selected. A combined muon is a muon reconstructed using information from both the inner detector and the muon spectrometer. The impact parameter resolution with respect to the primary vertex in data is shown in figure~\ref{fig:zmumu}. Only the inner detector component of the combined track is taken to extract the impact parameter distribution, and only tracks that have a measurement in the innermost pixel layer (B-layer) were considered for this comparison. The resolution in the core of the distribution is estimated by performing an iterative procedure to obtain the RMS within $3\sigma$ around the mean of the distribution. The NN clustering algorithm yields an improvement of 7\% (3\%) in the $z_0$ ($d_0$) resolution compared to the CCA clustering algorithm. This improvement varies significantly as a function of $\eta$, due to the variation in the incident angle of the tracks with respect to the module. At central $\eta$, no improvement in the $z_0$ resolution is obtained with the neural network algorithm because at normal incidence to the module the clusters are one pixel in size in the longitudinal direction and therefore interpolation is not possible.

\begin{figure}[htb!]
\begin{center}
\includegraphics[width=0.8\textwidth]{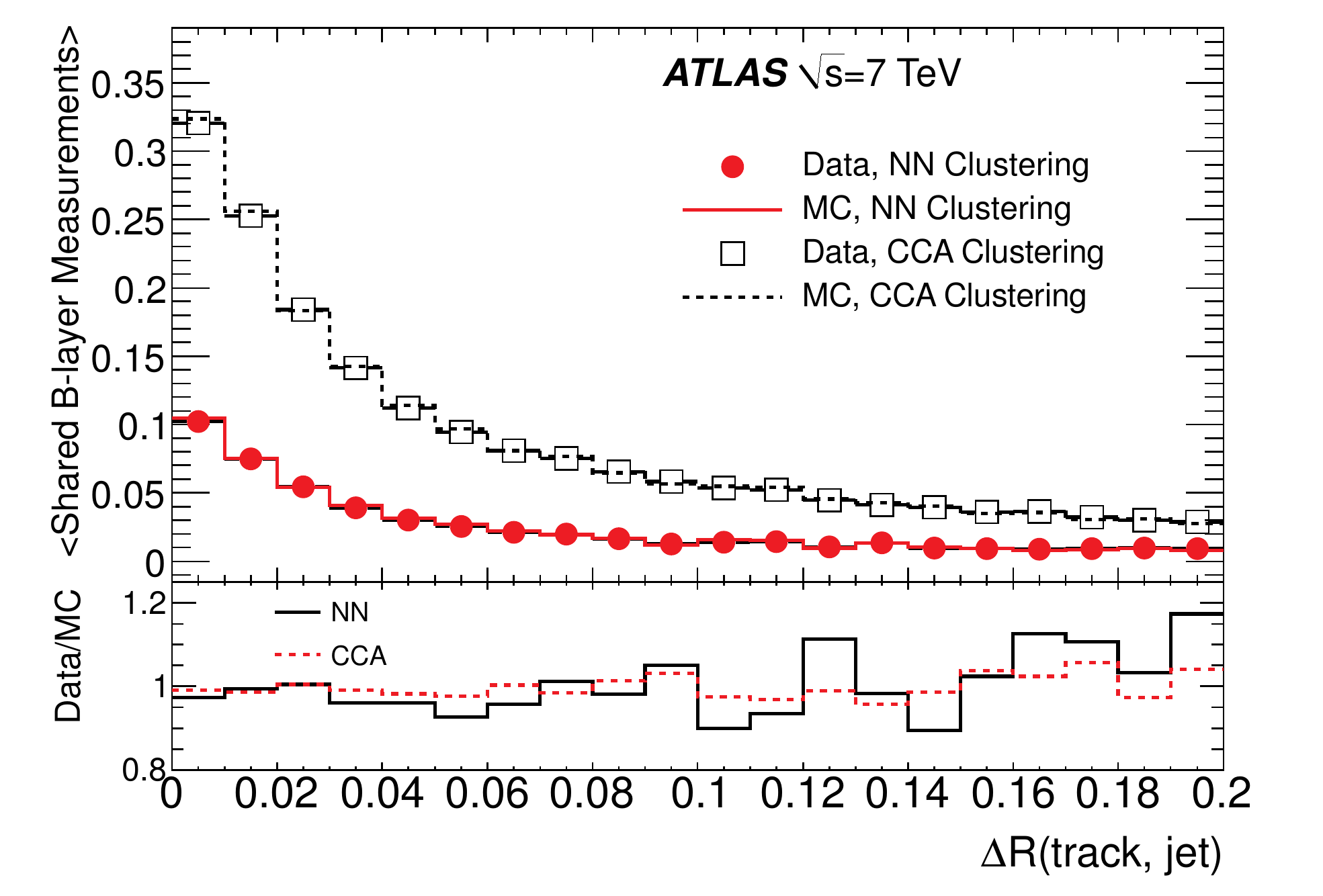}
\caption{The average number of shared measurements in the B-layer on tracks associated to anti--$k_{\mathrm{T}}$ jets with $500 < p_\mathrm{T} < 600$~GeV for data and simulation, reconstructed with the CCA and NN clustering algorithms. This is shown as a function of the distance of the track from the centre of the jet. The ratio of the average number of shared measurements in data and simulation is shown for both the NN (solid line) and the CCA (dashed line) clustering algorithms.
\label{fig:blayershared}}
\end{center}
\end{figure}

In addition to the improvement of the track resolution, the NN clustering algorithm suppresses clusters that are assigned to several track candidates. Figure \ref{fig:blayershared} shows the average number of shared measurements in the innermost pixel layer for tracks found in jets reconstructed from calorimetric information with $500 < p_\mathrm{T} < 600$~GeV, as a function of the distance $\Delta R$ to the centre of the jet. This distance is calculated from the angular separation as $\Delta R = \sqrt{\Delta \eta^2 + \Delta \phi^2}$. In the jet core, the average number of shared measurements is reduced by a factor of three. The simulation accurately describes the rate of shared measurements and the reduction obtained with the application of the neural network.

% heather check the pT cut

The NN clustering algorithm runs approximately six times slower than the CCA clustering algorithm. However, in comparison to the full event reconstruction, the NN splitting, the re--evaluation of the splitting during the track fitting and the increased combinations from additional track candidates increase the per--event execution time by at most 5 percent. This was estimated using the highest pile--up conditions experienced during normal physics data taking in 2012. 
 
\section{Conclusion}
\label{sec:conclusion}
A new method using a set of neural networks to identify and split clusters created by multiple charged particles in the ATLAS silicon pixel detector is presented. The algorithm results in a factor of three reduction of the numbers of measurements assigned to multiple tracks, in particular in the core of highly energetic jets.

An additional set of neural networks was trained to estimate cluster positions and uncertainties. The superior, non--linear behaviour of the neural network results in a significant improvement of the impact parameter resolution even for isolated tracks. Good agreement in neural network performance is observed when comparing Monte Carlo simulation and experimental data from proton--proton collisions at the LHC.

\section*{Acknowledgements}
We thank CERN for the very successful operation of the LHC, as well as the
support staff from our institutions without whom ATLAS could not be
operated efficiently.

We acknowledge the support of ANPCyT, Argentina; YerPhI, Armenia; ARC,
Australia; BMWF and FWF, Austria; ANAS, Azerbaijan; SSTC, Belarus; CNPq and FAPESP,
Brazil; NSERC, NRC and CFI, Canada; CERN; CONICYT, Chile; CAS, MOST and NSFC,
China; COLCIENCIAS, Colombia; MSMT CR, MPO CR and VSC CR, Czech Republic;
DNRF, DNSRC and Lundbeck Foundation, Denmark; EPLANET, ERC and NSRF, European Union;
IN2P3-CNRS, CEA-DSM/IRFU, France; GNSF, Georgia; BMBF, DFG, HGF, MPG and AvH
Foundation, Germany; GSRT and NSRF, Greece; ISF, MINERVA, GIF, I-CORE and Benoziyo Center,
Israel; INFN, Italy; MEXT and JSPS, Japan; CNRST, Morocco; FOM and NWO,
Netherlands; BRF and RCN, Norway; MNiSW and NCN, Poland; GRICES and FCT, Portugal; MNE/IFA, Romania; MES of Russia and ROSATOM, Russian Federation; JINR; MSTD,
Serbia; MSSR, Slovakia; ARRS and MIZ\v{S}, Slovenia; DST/NRF, South Africa;
MINECO, Spain; SRC and Wallenberg Foundation, Sweden; SER, SNSF and Cantons of
Bern and Geneva, Switzerland; NSC, Taiwan; TAEK, Turkey; STFC, the Royal
Society and Leverhulme Trust, United Kingdom; DOE and NSF, United States of
America.

The crucial computing support from all WLCG partners is acknowledged
gratefully, in particular from CERN and the ATLAS Tier-1 facilities at
TRIUMF (Canada), NDGF (Denmark, Norway, Sweden), CC-IN2P3 (France),
KIT/GridKA (Germany), INFN-CNAF (Italy), NL-T1 (Netherlands), PIC (Spain),
ASGC (Taiwan), RAL (UK) and BNL (USA) and in the Tier-2 facilities
worldwide.

\bibliographystyle{JHEP}
\bibliography{PCTFPaper}

\onecolumn 
\clearpage 
% ATLAS Collaboration author list
% Data extracted on 25-May-2014 for paper reference PERF-2012-05
%\documentclass[11pt]{article}
%\usepackage{a4wide}\begin{document}
\begin{flushleft}
{\Large The ATLAS Collaboration}

\bigskip

G.~Aad$^{\rm 84}$,
B.~Abbott$^{\rm 112}$,
J.~Abdallah$^{\rm 152}$,
S.~Abdel~Khalek$^{\rm 116}$,
O.~Abdinov$^{\rm 11}$,
R.~Aben$^{\rm 106}$,
B.~Abi$^{\rm 113}$,
M.~Abolins$^{\rm 89}$,
O.S.~AbouZeid$^{\rm 159}$,
H.~Abramowicz$^{\rm 154}$
% ...
\bigskip
\\
% ...
$^{11}$ Institute of Physics, Azerbaijan Academy of Sciences, Baku, Azerbaijan\\
% ...
$^{84}$ CPPM, Aix-Marseille Universit{\'e} and CNRS/IN2P3, Marseille, France\\
% ...
$^{89}$ Department of Physics and Astronomy, Michigan State University, East Lansing MI, United States of America\\
% ...
$^{106}$ Nikhef National Institute for Subatomic Physics and University of Amsterdam, Amsterdam, Netherlands\\
% ...
$^{112}$ Homer L. Dodge Department of Physics and Astronomy, University of Oklahoma, Norman OK, United States of America\\
$^{113}$ Department of Physics, Oklahoma State University, Stillwater OK, United States of America\\
% ...
$^{116}$ LAL, Universit{\'e} Paris-Sud and CNRS/IN2P3, Orsay, France\\
% ...
$^{152}$ Institute of Physics, Academia Sinica, Taipei, Taiwan\\
% ...
$^{154}$ Raymond and Beverly Sackler School of Physics and Astronomy, Tel Aviv University, Tel Aviv, Israel\\
% ...
$^{159}$ Department of Physics, University of Toronto, Toronto ON, Canada\\
% ...
\end{flushleft}

%\end{document}
% Created with xml2latex.py 
\end{document}